\documentclass[aps,pra,twocolumn,10pt,notitlepage]{revtex4-1}
\usepackage{amsmath}
\usepackage{braket}
\usepackage{color}
\usepackage{graphicx}
\usepackage{textcomp}
\usepackage{subcaption}
\usepackage{caption}
\usepackage{amssymb}
\usepackage{amsfonts}
\usepackage[parfill]{parskip}
\usepackage{hyperref}
\usepackage{multirow}
\usepackage{array}

\newcommand{\adag}[0]{a^{\dagger}}
\newcommand{\swbar}[0]{\ket{s_{\bar{w}}}}

\DeclareMathOperator{\csch}{csch}
\newcommand{\braswbar}[0]{\bra{s_{\bar{w}}}}

\usepackage[format=plain,justification=centerlast,singlelinecheck=true]{caption}
\begin{document}
  \title{Environment-assisted analog quantum search}
  \date{August 13, 2018}
\author{Leonardo Novo$^{1,2,3*}$}  
\author{Shantanav Chakraborty$^{1,2,3*}$}
\author{Masoud Mohseni$^4$}
\author{Yasser Omar$^{1,2}$}
\affiliation{$^1$Instituto de Telecomunica\c{c}\~oes, Physics of Information and Quantum Technologies Group, Lisbon, Portugal}
\affiliation{$^2$Instituto Superior T\'{e}cnico, Universidade de Lisboa, Portugal}
\affiliation{$^3$ Centre for Quantum Information and Communication, Ecole polytechnique de Bruxelles, CP 165, Universit\'{e} libre de Bruxelles, 1050 Brussels, Belgium
}
\affiliation{$^4$ Google Quantum AI Laboratory, Venice, California, USA}
\collaboration{$^*$both authors have equal contribution}

\begin{abstract}
Two main obstacles for observing quantum advantage in noisy intermediate-scale quantum computers (NISQ) are the finite precision effects due to control errors, or disorders, and decoherence effects due to thermal fluctuations. It has been shown that dissipative quantum computation is possible in presence of an idealized fully-engineered bath. However, it is not clear, in general, what performance can be achieved by NISQ when internal bath degrees of freedom are not controllable. In this work, we consider the task of quantum search of a marked node on a complete graph of $n$ nodes in the presence of both static disorder and non-zero coupling to an environment. We show that, given fixed and finite levels of disorder and thermal fluctuations, there is an optimal range of bath temperatures that can significantly improve the success probability of the algorithm. Remarkably for a fixed disorder strength $\sigma$, the system relaxation time decreases for higher temperatures within a robust range of parameters. In particular, we demonstrate that for strong disorder, the presence of a thermal bath increases the success probability from $1/(n \sigma^2)$ to at least $1/2$. While the asymptotic running time is approximately maintained, the need to repeat the algorithm many times and issues associated with unitary over-rotations can be avoided as the system relaxes to an absorbing steady state. Furthermore, we discuss for what regimes of disorder and bath parameters quantum speedup is possible and mention conditions for which similar phenomena can be observed in more general families of graphs. Our work highlights that in the presence of static disorder, even non-engineered environmental interactions can be beneficial for a quantum algorithm.
\end{abstract}
\maketitle
\section{Introduction}
A major obstacle to the development of a scalable quantum computer is its interaction with an environment, resulting in decoherence and loss of quantum advantage \cite{divincenzo_criteria, palma_dissipation}. Even if a quantum system is well isolated from the environment, there are always experimental imperfections in the setting of the system's parameters which can lead to a unitary dynamics different from the desired one and thus to errors in the quantum computation. In the circuit model, these sources of error can be countered using various error correction techniques \cite{nielchuang}. However, these have proven to be rather expensive as they require a huge overhead in terms of the number of qubits \cite{fowler_surface_code}. Furthermore, for alternative models of quantum computation such as adiabatic \cite{farhi_adiabatic, aharonov2008adiabatic} or quantum walks \cite{childs_universal,childs2013universal}, the theory of error correction is much less developed or nonexistent \cite{jordan2006error,young2013error1,young2013error2}. Dissipative quantum computation has also been proposed \cite{verstraete2009quantum}, however, the necessary system-environment interaction must be highly engineered, which is extremely challenging.  

On the other hand, there are quantum processes that are enhanced by naturally occuring interactions with an external environment. It has been shown that quantum transport in certain disordered structures like protein complexes in biological systems \cite{mohseni2008environment,plenio2008dephasing,
rebentrost2009environment,caruso2009highly,mohseni2014quantum} and others \cite{wu2013generic,novo2016disorder} can be enhanced for certain ranges of environment parameters. A simplified interpretation of this behavior is that in a disordered quantum system there are destructive interferences suppressing quantum transport \cite{anderson1958absence} and since decoherence processes suppress these destructive interferences, transport efficiency is enhanced \cite{lloyd2011quantum}. Also, relaxation dynamics coming from the interaction with a thermal bath can significantly improve quantum transport provided that the bath spectral density is in a regime which enhances certain desired transitions \cite{rey2013exploiting}.

In this article, we explore whether a non-engineered environment can benefit a quantum algorithm. We address this by considering the analog version of Grover's algorithm \cite{Farhi_analog_grover} which can be seen as an instance of search by CTQW on the complete graph of $n$ nodes \cite{Childs_spatial_search}. This algorithm finds a node in the graph, which is marked by an oracular Hamiltonian, starting from an equal superposition of all the nodes of the graph, in $\mathcal{O}(\sqrt{n})$ time. This running time is quadratically faster than the best known classical algorithm,  and  is optimal \cite{Farhi_analog_grover}.\\
We consider the effect of a static diagonal disorder term of strength at most $\sigma$ in the search Hamiltonian which can be interpreted as a faulty oracle. We show that for $\sigma>\mathcal{O}(1/\sqrt{n})$ the algorithm loses its optimality. Above this threshold, we find that the maximum probability of success decreases with the size of the system and several repetitions are needed to find the marked node. 

By coupling the system to a thermal environment \cite{leggett1987dynamics,gardiner2004quantum,breuer}, the transition from the initial state to the marked node, which was suppressed in the unitary case due to disorder, is now enhanced because of thermal relaxation. This is because the dynamics occurs mostly in a two-dimensional subspace spanned by the ground and first excited states of the system, where the ground state has a large overlap with the marked node. So, the system relaxes to a thermal state which has a constant overlap with the solution and hence the algorithm exhibits a fixed-point property. Thus only a constant number of repetitions are needed to find the marked node and a measurement can be made at any time after the system relaxes. Interestingly, the relaxation time and thus the algorithmic running time improves with temperature as long as the two-level approximation is valid. For the maximum allowed temperature and for a fixed disorder strength $\sigma$, the scaling of this relaxation time matches the running time of the corresponding closed system with the same disorder strength, up to logarithmic factors.

Our work contrasts with the idea of engineering the dissipation of a quantum system in order to drive a quantum computation \cite{verstraete2009quantum}. Instead, we study how a naturally occuring coupling to a thermal bath can help when static errors are present in the system Hamiltonian. This way, our results also differ from those concerning thermal effects in adiabatic quantum computation~\cite{childs2001robustness, amin2008thermally,devega2010,albash2012quantum,dickson2013thermally,wild2016}. 

Before proceeding with a careful analysis of the scaling of the running time with the different bath parameters, let us look at the closed system behavior in the presence of static disorder and analyze the algorithm in that scenario.

\section{Analog quantum search with diagonal disorder.}\label{sec:search_dis}
Let $G$ be a graph with $n$ vertices $V=\{1,2,..,n\}$. We consider the Hilbert space spanned by the localized quantum states at the vertices of the graph $\mathcal{H}=\text{span}\{\ket{1},...\ket{n}\}$ and the search Hamiltonian given by
\begin{equation}
\label{eqmain:search_ham}
H_{search}=-\ket{w}\bra{w}-\gamma A_G,
\end{equation}
where $\ket{w}$ corresponds to the solution of the search problem, $\gamma$ is a real number and $A_G$ is the adjacency matrix of graph $G$ \cite{Childs_spatial_search}. The algorithm is said to be optimal on graph $G$ if starting from the equal superposition of all states, i.e.\ $\ket{s}=\sum_{i=1}^n \ket{i}/\sqrt{n}$, there is a value of $\gamma$ such that the probability of finding the solution node $\ket{w}$ upon a measurement in the vertex basis after a time $T=\mathcal{O}(\sqrt{n})$ is constant, irrespective of $w$. Here we consider quantum walk on a complete graph which is equivalent to the analog quantum search algorithm introduced in \cite{Farhi_analog_grover}. The search Hamiltonian in that case is given by
\begin{equation}
\label{eqmain:search_ham_analog}
H_{search}=-\ket{w}\bra{w}-\ket{s}\bra{s},
\end{equation}
where we have chosen $\gamma=1/n$. The gap between the ground state and the first excited state, up to an error of $\mathcal{O}(1/n)$ is $\Delta=2/\sqrt{n}$. The dynamics of the algorithm is a rotation in a two-dimensional subspace containing the initial state $\ket{s}$ and $\ket{w}$. The success probability $P_w(t)=\sin^2(t/\sqrt{n})$ is close to one after a time $T=\pi\sqrt{n}/2$.

The analog search algorithm requires an oracle that marks the solution node to an energy that is different from the rest of the nodes. In order for the problem to have a fair comparison to the standard Grover's algorithm in the circuit model, the energy at the marked node is chosen to be $-1$ \footnote{This energy at the marked node implies that the quantum simulation of $\ket{w}\bra{w}$ for time $t$ would correspond to $\mathcal{O}(t)$ queries to the standard Grover oracle}. However, an imperfect implementation of the oracle  might severely affect the algorithmic performance. We define an imperfect oracle as one which ``marks" each node of the graph erroneously: each non-solution node $j$ is marked with an energy $\epsilon_j$, while the solution node, $w$ is marked with an energy $-1+\epsilon_w$ (where each $\epsilon_w$ is a random variable). 
The resultant effect can be perceived as static disorder on the nodes of the complete graph. Furthermore we assume that these errors occurring due to imperfect implementations are fixed in nature, i.e.\ each $\epsilon_i$ remains fixed across multiple iterations of the algorithm. The case where the instance of oracular defect varies over iterations has been discussed in Ref.~\cite{shenvi2003effects}. In our case, we have the following search Hamiltonian
\begin{equation}
\label{eqmain:ham_dis}
H^{dis}_{search}=-\ket{w}\bra{w}-\ket{s}\bra{s}+\sum_{i=1}^{n}\epsilon_i\ket{i}\bra{i},
\end{equation}
where $\epsilon_i$-s are the value of static disorder at vertex $i$ and are i.i.d random variables from some probability distribution of mean $0$ and standard deviation $ \sigma \ll 1 $. In fact, the form of the probability distribution is not very important for the results we derive, as long as there is a high probability that $|\epsilon_i|<\sigma$, and also that in a typical instance we have $\epsilon_i$ to be of the same order as $\sigma$. 

\begin{figure}[h!]
\centering
\includegraphics[scale=0.45]{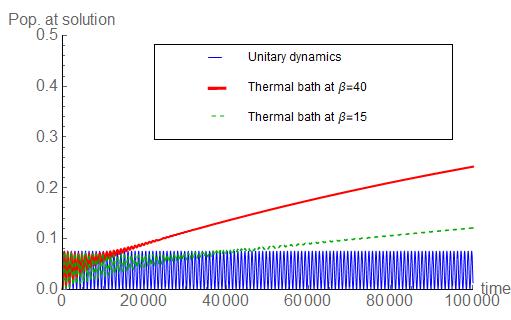}
\caption{Population at the solution node versus time for a complete graph of $10^6$ nodes where each node of the graph is affected by disorder of maximum strength $\sigma=0.007$ for the cases of no bath (oscillatory thin blue curve), thermal bath at inverse temperature $\beta=40$ (thick red curve) and at $\beta=15$ (dashed green curve). We find that the population at the solution is always low in the unitary regime implying that the algorithm needs to be repeated several times. In the case where the disordered system is coupled to a thermal bath, we consider that the bath has an ohmic spectral density with a cut-off frequency $\omega_c=2$ and system-bath coupling $g=0.02$. We numerically solve the Bloch-Redfield master equation. The interaction with the thermal bath results in amplifying the population at the solution with time without compromising much in the algorithmic running time. Moreover, increasing the temperature of the bath ensures faster relaxation and improves the running time of the algorithm. }
\label{fig:pop_sol_weak_dis_0_temp_open_vs_closed_system_main}
\end{figure}

The approximate eigenstates and eigenvalues of $H_{\text{search}}^\text{dis}$ are calculated in Appendix~\ref{sec:analog_search_static_errors}, whereas here we summarize the results. Let $\swbar$ be the equal superposition of all nodes other than the solution node $\ket{w}$. Then by using degenerate perturbation theory, we find that the approximate ground and first excited states of the system are obtained by diagonalizing the search Hamiltonian projected onto the subspace spanned by $\{\ket{w}, \swbar\}$. The Hamiltonian of the effective two level system is 
\begin{equation}\label{eqmain:projectedham}
H_{red}=\begin{bmatrix}
-1+\epsilon_w && -1/\sqrt{n}\\
-1/\sqrt{n} && -1
\end{bmatrix},
\end{equation}
which interestingly only depends on the error at the oracle $\epsilon_w$. The gap between the ground state and the first excited state of the perturbed Hamiltonian is \begin{equation}
\Delta\approx\sqrt{\epsilon_w^2+4/n}
\end{equation} 
and the success probability of the algorithm is given by 
\begin{equation}
\label{eqmain:succ_prob_disorder}
P_w(t)=|\braket{w|e^{-iH_{red} t}|s}|^2\approx\frac{1}{1+n\epsilon_w^2/4}\sin^2(\frac{\Delta t}{2}),
\end{equation}
which is plotted in blue in Fig.~\ref{fig:pop_sol_weak_dis_0_temp_open_vs_closed_system_main}.                
The maximum success probability is achieved at a time $T=\pi/\Delta$ and since it is lower than 1, we need to repeat the algorithm $1/P_w(T)$ times on average to find the marked node. Hence, Eq.~\eqref{eqmain:succ_prob_disorder} shows that there are two distinct regimes for the average running time

$\bullet$ \textit{Weak disorder} $\left(\sigma\leq \mathcal{O}\left(1/\sqrt{n}\right)\right)$:~The maximum success probability is constant and the frequency $\Delta=\mathcal{O}(1/\sqrt{n})$. Thus, the algorithm remains optimal.
~\\ 
$\bullet$ \textit{Strong disorder} $\left(\sigma>\mathcal{O}\left(1/\sqrt{n}\right)\right)$:~The maximum success probability scales as $\mathcal{O}(1/(n\sigma^2))$ and $\Delta=\mathcal{O}(\sigma)$. Thus, one needs to repeat the algorithm $\sim n\sigma^2$ times on average, to obtain an expected running time of $\mathcal{O}(n\sigma)$.

This implies that a high degree of control in the system is necessary to maintain quantum speed-up. In fact, unless it is possible to decrease the disorder strength $\sigma$ with the system size, only a constant speed-up is possible with respect to the classical case where search takes~$\mathcal{O}(n)$ time. We note, however, that for classical unstructured search the average running time is n/2 whereas we can have $\sigma\ll 1$ if we have good control over the quantum system.

\section{Analog quantum search in the presence of a thermal bath}
We shall now see how the coupling of the system to a thermal bath can increase the success probability of the algorithm due to thermal relaxation. We will focus our analysis on the strong disorder regime, since it is more realistic to assume that we would not have sufficient control on the system to ensure that the disorder strength $\sigma$ is less than $1/\sqrt{n}$, given that the dimension of the Hilbert space $n$ increases exponentially with the number of qubits. The weak disorder regime is treated in Appendix \ref{appendix_D}.                  

By looking at the approximate two level description of $H^{dis}_{search}$ given in Eq.~\ref{eqmain:projectedham} one can see that the transitions from $\swbar$ to $\ket{w}$ are suppressed due to the energy mismatch $\epsilon_w$. On coupling the system to a thermal bath, we expect it to evolve to a thermal state which enhances the aforementioned transition due to thermal relaxation. In fact, in the zero temperature regime, we expect the system to relax to the ground state and thus, if the ground state has a large overlap with $\ket{w}$, we obtain a maximum probability of success close to 1, in spite of disorder.

In the strong disorder regime, we obtain that the ground state of $H_{red}$ is approximately $\ket{w}$ only if the random variable $\epsilon_w\ll -1/\sqrt{n}$, which happens with probability of approximately $1/2$ assuming that the probability distribution is symmetric around $0$.  In order to ensure that the state $\ket{w}$ almost always has a large overlap with the ground state we choose the parameter $\gamma=(1-\sigma)/n$, instead of the value $1/n$ mentioned before and chosen in Ref.~\cite{Farhi_analog_grover}. This choice does not change the scaling of the average running time of the search algorithm with disorder, which is still $\mathcal{O}(n\sigma)$ on average and requires $\sim n\sigma^2$ repetitions. The gap between the ground state and the first excited state, as result of this choice of $\gamma$ becomes
$\Delta=\sigma-\epsilon_w+\mathcal{O}(1/(n\sigma))$ and the approximate eigenstates to first order in perturbation theory are
\begin{align}
\label{eqmain:approx_eigenstates_disorder1}
\ket{\lambda_{1}}&\approx \ket{w}+\frac{1}{\sqrt{n} (\sigma-\epsilon_w)}\ket{s_{\bar{w}}},\\
\label{eqmain:approx_eigenstates_disorder2}
\ket{\lambda_{2}}&\approx \frac{1}{\sqrt{n} (\sigma-\epsilon_w)}\ket{w}-\ket{s_{\bar{w}}},
\end{align}
which are obtained in an analogous way as shown in Appendix \ref{sec:analog_search_static_errors}. 
With this new choice of $\gamma$, the overlap of the ground state with the marked node is close to 1 with high probability, as desired.

We consider the following Hamiltonian which describes the interaction of the system with a thermal bath of harmonic oscillators
\begin{equation}
\label{eqmain:interaction_hamiltonian}
H_I=\sum_{i=1}^n\sum_{\alpha}g_{i\alpha}(a_{i\alpha}+\adag_{i\alpha})\ket{i}\bra{i},
\end{equation}
where $\adag_{i\alpha}$ and $a_{i\alpha}$ are the bath creation and annihilation operators obeying $[a_{i\alpha},\adag_{j \beta}]=\delta_{i,j}\delta_{\alpha,\beta}$, i.e.\ we consider that each node $\ket{i}$ of the complete graph is coupled to a bosonic bath, which we assume to be at an inverse temperature $\beta$ (throughout the article we are working in units where Boltzmann constant $k_B=1$). Furthermore we assume that the bath temperature is low enough so that the transitions to states higher than the first excited state are negligible. To ensure that this happens we need $\beta\gg\beta^*=\mathcal{O}\left(\log(n)\right)$. 

To describe the evolution of the system's density matrix we first assume that the coupling between each site of the system and the bath is considered to be identical ($g_{i\alpha}=g,~\text{for all} ~i,\alpha$) and that $g$ is sufficiently weak so that the system and the bath remain uncorrelated at all times. Secondly, we consider that the time scale of decay of the bath correlation functions $\delta t$ is much faster than the relevant time-scales of the system, i.e.\ the Markov approximation is valid. The condition $g\ll 1/\delta t$ ensures that this is indeed the case. These assumptions lead us to the well known Bloch-Redfield master equation \cite{bookblum,gardiner2004quantum,breuer}. This equation allows us to resolve system time-scales which, for the weak disorder regime, are of $O(\sqrt{n})$ and thus are important to understand the regimes where the algorithm remains optimal. This analysis is done in Appendix~\ref{sec:bloch_redfield_two_level},~Appendix~\ref{appendix_C} and Appendix~\ref{appendix_D}. In the strong disorder regime, we are not interested in resolving this the system time-scales, which are of $O(1/\sigma)\ll \sqrt{n}$, and so we can take the secular approximation \cite{breuer}. The condition $g\ll\sqrt{\sigma/\delta t}$ ensures that both the secular and Markov  approximations are valid (see Appendix~\ref{appendix:markov_and_secular_approximation}). 
 
Let $\rho_{i j}=\bra{\lambda_i}\rho\ket{\lambda_j}$ be the density matrix elements of the system, expressed in its eigenbasis. The master equation that describes the time-evolution of the population of ground and first excited states of the system after taking the secular approximation is
\begin{align}
\label{eqmain:master_eq_pop_secular}
\dot{\rho}_{kk}&= \sum_{l\neq k}W_{kl}\rho_{ll}-\sum_{l\neq k}W_{lk}\rho_{kk},
\end{align}
where $k\in\{1,2\}$ and $l\in\{1,2\}$. The transition rates are given by
\begin{align}
\label{eqmain:transition_rates_pop_secular}
W_{kl}=\begin{cases}
2\pi J(\omega_{kl})\Lambda_{kl}\mathcal{N}(\omega_{kl}),\text{~~~~~~~~~~~~~~~~~~} \omega_{kl}\geq 0\\
2\pi J(\omega_{lk})\Lambda_{kl}\Big{(}\mathcal{N}(\omega_{lk})+1\Big{)},\text{~~~~~~~~~} \omega_{kl}< 0
\end{cases}
\end{align} 
with $\mathcal{N}(\omega)=1/(e^{\beta\omega}-1)$ and $\Lambda_{kl}=\sum_{i}|c_{ik}c_{il}|^2$. The coefficients $c_{ik}$ are obtained from the basis change $\ket{i}=\sum_{k}c_{ik}\ket{\lambda_k}$. We consider the spectral density of the bath to be ohmic with an exponential cut-off, i.e.\ $J(\omega)=\eta g^2\omega e^{-\omega/\omega_c}$, and that the cut-off frequency $\omega_c $ to be a constant larger than the system energy scale, i.e. $\omega_c>1$. Also, $\eta$ is a constant normalization factor.

From Eq.~\eqref{eqmain:approx_eigenstates_disorder1}, we see that the population at the solution is approximately the population of the ground state. Using this and Eq.~\eqref{eqmain:master_eq_pop_secular}, we obtain  
\begin{align}
\label{sol_diff_eq}
P_w(t)\approx&\rho_{11}(t)+\mathcal{O}\left(\dfrac{1}{\sigma\sqrt{n}}\right)\\
\approx&\dfrac{1-e^{-t/T_{rel}}}{1+e^{-\beta\Delta}}+\mathcal{O}\left(\dfrac{1}{\sigma\sqrt{n}}\right).
\end{align}
The relaxation time is
\begin{equation}
\label{eqmain:relaxation_time_general_expression}
T_{rel}\sim\dfrac{1}{\Lambda_{12} J(\Delta)}\tanh\left(\dfrac{\beta\Delta}{2}\right),
\end{equation}
where $\Lambda_{12}$ can be calculated from Eqs.~\eqref{eqmain:approx_eigenstates_disorder1}~and \eqref{eqmain:approx_eigenstates_disorder2} which yield $\Lambda_{12}=\mathcal{O}\left(1/(n\sigma^2)\right)$. We obtain thus a quantum algorithm that is run simply by thermal relaxation and whose running time is given by $T_{rel}$. The probability of success is given by the ground state population of the Gibbs state 
\begin{equation}
P_{suc}=(1+\exp(-\beta \Delta))^{-1}
\end{equation}
which is always larger than $1/2$. This is an important advantage with respect to the unitary, disordered algorithm since the population at the solution node only increases with time and the probability of success is much larger. This way, only two or less repetitions or the algorithm are needed, on average, to find the marked node in contrast with the $O(n \sigma^2)$ number of repetitions needed on average in the disordered unitary case (see Eq.~\eqref{eqmain:succ_prob_disorder}). However, a careful analysis of the relaxation time is needed to ensure that any quantum advantage remains.

\textit{Zero temperature ($\beta\rightarrow\infty$):} When the thermal bath is at zero temperature, i.e.\ when $\beta\rightarrow\infty$, the relaxation time of the system is $T_{rel}(\infty)=\mathcal{O}\left(n\sigma/\eta g^2\right)$.

\textit{High temperature $\left(\beta^*\ll\beta\ll 1/\sigma\right)$:} In this regime of temperature, $\tanh(\beta\sigma/2)\approx\beta\sigma/2$. This gives us that the relaxation time, $T_{rel}(\beta)=\mathcal{O}\left(n\sigma^2\beta/\eta g^2\right)$. Thus the ratio,
\begin{equation}
\tau=\dfrac{T_{rel}(\beta)}{T_{rel}(\infty)}=\beta\sigma\ll 1.
\end{equation}
This shows that increasing the temperature actually ensures faster relaxation to the thermal state thereby improving the algorithmic running time. This has been plotted in Fig.~\ref{fig:pop_sol_weak_dis_0_temp_open_vs_closed_system_main} where we find that relaxation is faster for the thermal bath at $\beta=15$ (green) as compared to $\beta=40$ (red). Also observe the difference in the dynamics of the population at solution of these two curves as compared to the unitary scenario (blue). The probability at the solution is considerably higher in the presence of a thermal bath.

In order to analyse the fastest relaxation time we can obtain in this framework, it is crucial to note that the validity of the secular and Markov approximations implies that we have to restrict the system-bath coupling to a value $g\ll\sqrt{\sigma/\delta t}$. The larger the $g$ the faster the relaxation, and so the relaxation time is minimized for $g=\chi \sqrt{\sigma/\delta t}$, where $\chi$ is some small constant.  

We prove in Appendix~\ref{sec:bath_correlation_function_zero_temp} and Appendix \ref{appendix:bath_correlation_function_non_zero_temp} that the bath-correlation timescale $\delta t$ is $\delta t\sim\omega_c$ at zero temperature and is given by $\delta t\sim\beta$ at finite temperature. This implies, for zero temperature the lower bound for the relaxation time is $T_{rel}(\infty)=\Omega\left(n\right)$ which is no better than classical search. For finite temperatures however, we have that $T_{rel}(\beta)=\Omega\left(n\sigma\beta^2\right)$. The relaxation time decreases for higher temperatures but it is necessary to keep $\beta>\mathcal{O}\left(\log(n)\right)$ for the two-level approximation to be valid. Hence, the fastest relaxation possible in this framework is
\begin{equation}
T_{rel}=\mathcal{O}\left(n\sigma(\log{n})^2\right),
\end{equation} 
which matches the running time of the unitary disordered case up to a logarithmic factor.

We have thus demonstrated that the success probability of the algorithm improves drastically in the presence of the bath as compared to the disordered unitary quantum algorithm, despite a small (logarithmic) overhead in terms of asymptotic running time. We leave as an open question whether also an asymptotic improvement in running time can be achieved for other models of the bath and system-bath interaction. For completeness, we show in Appendix~\ref{sec:lower_bound_search_environment}, by a simple adaptation of the proof of Ref.~\citep{Farhi_analog_grover}, that the lower bound for any quantum search algorithm interacting with an external system is $\mathcal{O}(\sqrt{n})$. Given the model we considered, we can show that this bound is attained for the case of weak disorder and zero temperature, as demonstrated in Appendix~\ref{appendix_C}.
\section{Environment-assisted quantum search on other graphs}
\label{sec:env_assisted_other_graphs}
The results derived so far have concerned the problem of searching a marked node in a complete graph. However, the same effects are expected to happen for quantum search on graphs whose adjacency matrix has a large spectral gap, such that the dynamics of the search problem happens mostly in a two-dimensional subspace.

In Ref.~\cite{almost_all}, the authors show a sufficient condition for the spatial search algorithm to be optimal on any graph $G$, the spectrum of the normalized adjacency matrix of $G$, $A_G$, should satisfy the following properties: (i) the gap between two highest eigenvalues of $A_G$ is constant and (ii) the overlap of $\ket{w}$ with the eigenstate $\ket{v_1}$, corresponding to the largest eigenvalue of $A_G$, is $\mathcal{O}\left(1/\sqrt{n}\right)$. Then starting from $\ket{v_1}$, the algorithm evolves to a state close to $\ket{w}$ in $\mathcal{O}(\sqrt{n})$ time. Several classes of graphs obey this sufficient condition such as Erd\"os-Renyi random graphs which are graphs of $n$ nodes such that each edge exists between any two of these nodes with probability $p$, random regular graphs \cite{almost_all}, strongly regular graphs \cite{Meyer_symmetry} or complete bipartite graphs \cite{dimred}.

In these cases, the search Hamiltonian, similarly to the case of the complete graph, has its ground and first excited state with energies $-1\pm \Delta/2$, where $\Delta$ is of $\mathcal{O}(1/\sqrt{n})$. Moreover, the energy gap between the ground state and second excited state $\tilde{\Delta}=\lambda_3-\lambda_1$, is much larger than $\Delta$. In this situation, if we consider the problem with static disorder and we have $\sigma\ll \tilde{\Delta}$, the perturbation theory arguments applied in Sec.~\ref{sec:search_dis} hold and the success probability of the algorithm should decrease drastically for $\sigma>\mathcal{O}(1/\sqrt{n})$. Furthermore, the coupling of the system to a thermal bath of inverse temperature $\beta\gg\log{(n)}/\tilde{\Delta}$ will induce thermal relaxation and increase the probability at the marked node while maintaining the population at the higher excited states negligible.
We expect thus that environment-assisted effects on quantum search algorithms by quantum walk to happen for several classes of graphs. It would be interesting to show also, whether such effects hold for quantum search on graphs whose topology change with time such as the random temporal networks \cite{chakraborty2017optimal}, or for quantum state transfer protocols based on quantum search \cite{almost_all}. 

\section{Discussion} 
We have analyzed the robustness of quantum analog search algorithm in the presence of diagonal static disorder and showed that the algorithm loses optimality for a disorder strength $\sigma>\mathcal{O}(1/\sqrt{n})$. In this regime, the success probability decreases with the system size and the algorithm needs to be repeated $n \sigma^2$ times on average, to have a running time of $ \mathcal{O}{(n \sigma)}$. 

We have shown that, if this system is coupled to a thermal bath, it is possible to significantly increase the success probability of the algorithm, from $1/(n\sigma^2)$ to a fixed value larger than $1/2$, due to thermal relaxation. Moreover, the algorithmic running time improves with temperature due to faster relaxation. For an appropriate choice of bath parameters, we obtain an algorithm in the open regime whose running time is close to that of the disordered unitary case, with the added advantage that only a constant number of repetitions are needed. Similar effects are possible for search by continuous time quantum walk on graphs whose adjacency matrix has a large spectral gap, which includes random graphs \cite{almost_all}.

It is important to point out the contrast between our result and the previous studies of environment-assisted quantum transport. The known results on excitonic transport study mostly the efficiency of transport towards a trapping site and how it improves by coupling the system to an environment. The modelling of the trapping process is done by a non-unitary term acting locally at this site. Also, these studies focus on small system sizes, that model light-harvesting molecules. On the other hand, for the quantum search problem the aim is to calculate the time needed for the wave-function to localize at a certain marked node of a class of graphs and, most importantly, how this time scales with the the problem size, which is assumed to be large. Also, the solution of the search problem is marked by a local Hamiltonian term, which affects the unitary evolution of the system. Thus, both the figures of merit and the equations of motion describing the dynamics are different when analysing the problem of quantum transport to a trapping site and quantum search of a marked node.

In the context of quantum computation, the question of how different error parameters should be controlled in order to obtain a certain quantum advantage is crucial for near-term non-error corrected quantum devices \cite{preskill2018quantum}. Such studies can guide experimentalists when scaling up these systems. Our work highlights the importance of  controlling the strength of the static errors $\sigma$. For the search problem, in both the unitary and non-unitary scenario, this parameter dictates how much advantage we have with respect to classical search. On increasing the dimension of the search space, to maintain a certain quantum advantage, it is necessary to decrease the value of $\sigma$ accordingly. It is interesting to note, that in the non-unitary case, such control is not required for temperature. In fact, the algorithm performs better for larger temperatures, due to faster thermal relaxation, and the only restriction is that $\beta>\mathcal{O}(\log(n))$. 

Our work can be extended in several ways. It would be interesting to explore whether a similar environment-assisted effects holds when there is an unkown number of solutions to the search problem. In such a case, the dissipative dynamics could lead to a new quantum algorithm for fixed-point search assisted by the environment, requiring no additional resources \cite{yoder2014fixed,dalzell2017fixed}. Furthermore, it would be worth exploring whether other models for bath and system-bath interaction, possibly taking into account non-Markovian effects, could lead to faster thermal relaxation and whether it is possible to get close to the proven lower bound of $\mathcal{O}(\sqrt{n})$ even in the presence of strong disorder \cite{rey2013exploiting, de2017dynamics}. 

Finally, our work suggests that naturally occurring open quantum system dynamics can be advantageous for analog algorithms affected by static errors.
\\
\begin{acknowledgments}
\textit{Acknowledgments---}LN, SC and YO thank the support from Funda\c{c}\~{a}o para a Ci\^{e}ncia e a Tecnologia (Portugal), namely through programmes PTDC/POPH/POCH and projects UID/EEA/50008/2013, IT/QuNet, ProQuNet, partially funded by EU FEDER, and from the JTF project NQuN (ID 60478).~Furthermore, LN and SC acknowledge the support from the DP-PMI and FCT (Portugal) through scholarships SFRH/BD/52241/2013 and SFRH/BD/52246/2013 respectively. LN also acknowledges the support from the Fondation Wiener-Anspach. MM acknowledges support from Quantum Artificial Intelligence Laboratory at Google.~SC also acknowledges the hospitality of QuSoft, CWI Amsterdam where a part of the work was done.
\\
\end{acknowledgments}
\onecolumngrid
\hrulefill
\appendix
\appendix
\label{appendix}
\section{Analog quantum search with static errors}
\label{sec:analog_search_static_errors}
The analog search algorithm requires an oracle that marks the solution node to an energy that is different from the rest of the nodes. In order for the problem to have a fair comparison with the standard Grover's algorithm in the circuit model, the energy at the marked node is chosen to be $-1$ \citep{Childs_spatial_search}. However an imperfect implementation of the oracle  might severely affect the algorithmic performance. We define an imperfect oracle as one which ``marks" each node of the graph erroneously: each node non-solution node $j$ is marked with an energy $\epsilon_j$, while the solution node, $w$ is marked with an energy $-1+\epsilon_w$ (where each $\epsilon_w$ is a random variable). The resultant effect can be perceived as an introduction of static disorder to the nodes of the complete graph. We consider that these errors are systematic i.e.\ we assume that the value of each $\epsilon_j$ does not change over different iterations of the algorithm. We have thus the following search Hamiltonian
\begin{equation}
\label{eq:ham_dis}
H^{dis}_{search}=-\ket{w}\bra{w}-\ket{s}\bra{s}+\sum_{i=1}^{n}\epsilon_i\ket{i}\bra{i},
\end{equation}
where $\epsilon_i$ is the value of static disorder at vertex $i$ and are i.i.d random variables from some probability distribution of mean $0$ and width $ 2 \sigma$. In fact, the form of the probability distribution is not very important for the results we derive, as long as there is a high probability that $-\sigma\leq\epsilon_i\leq\sigma$, and also that in a typical instance we have $\epsilon_i$ to be of the same order as $\sigma$. We assume that $\sigma\ll 1$ and that one can estimate the value of $\sigma$ without having access to the individual $\epsilon_i$s. Using perturbation theory we calculate the approximate system eigenstates.
To do so, let us rewrite the Hamiltonian in the following form
\begin{equation}
\label{ham_dis}
H^{dis}_{search}=\underbrace{-\ket{w}\bra{w}-\swbar\braswbar}_{H_0}-\underbrace{\frac{1}{\sqrt{n}}(\ket{w}\braswbar+\swbar\bra{w})+\sum_{i=1}^{n}\epsilon_i\ket{i}\bra{i}}_V,
\end{equation}
where we have neglected terms of order $O(1/n)$. It is expected that the strength of the perturbation $V$ should be dominated by the disorder when $\sigma\gg 1/\sqrt{n}$, whereas if $\sigma\ll 1/\sqrt{n}$ we expect the algorithm to be unaffected by disorder. In fact we will see that this threshold is very important for the running time of the algorithm.

Since $H_0$ has two degenerate eigenstates $\ket{w}$ and $\swbar$, we apply degenerate perturbation theory to obtain the approximate ground and first excited states of the system. At first order, these are calculated by the diagonalization of the Hamiltonian projected onto this degenerate subspace, which is given by
\begin{equation}\label{eq:projectedham}
H_{red}=\begin{bmatrix}
-1+\epsilon_w && -1/\sqrt{n}\\
-1/\sqrt{n} && -1+\bar{\epsilon}
\end{bmatrix},
\end{equation} 
where, $\epsilon_w$ is the strength of disorder at the solution node $\ket{w}$ and $\bar{\epsilon}=\sum_{i\neq w}\epsilon_i/(n-1)$ is the mean of the disorder at all sites other than the solution. We will neglect the random variable $\bar{\epsilon}$ because it has 0 mean and its fluctuations are of $\mathcal{O}(\sigma/\sqrt{n})$. This is smaller than the other perturbation term, $\epsilon_w$ which is $\mathcal{O}(\sigma)$. In any case, neglecting $\overline{\epsilon}$ will affect the success probability by a relative error of $\mathcal{O}(\sigma)$. 

From Eq.~\eqref{eq:projectedham} it is clear that the dynamics is dominated by the value of disorder at the marked vertex. The diagonalization of $H_{red}$ yields the following eigenvectors 
\begin{align}
\ket{\lambda_1^{(1)}}&=\frac{1}{K}\left(\dfrac{1}{\sqrt{n}}\ket{w}+\left(\Delta-\frac{\epsilon_w}{2}\right)\swbar\right)\\
\ket{\lambda_2^{(1)}}&=\frac{1}{K}\left(\left(\dfrac{\epsilon_w}{2}-\Delta\right)\ket{w}+\frac{1}{\sqrt{n}}\swbar\right),
\end{align}
where $K=\sqrt{(\frac{\epsilon_w}{2}-\Delta)^2+1/n}$ is the normalization factor. The corresponding eigenvalues are
\begin{align}
\lambda_{1}^{(1)}&=-1+\epsilon_w/2-\Delta\\
\lambda_{2}^{(1)}&=-1+\epsilon_w/2+\Delta
\end{align}
where the gap $\Delta$ is given by $\Delta=\lambda_2-\lambda_1=\sqrt{\epsilon_w^2+4/n}$. The success probability of the algorithm, also calculated at first order in perturbation theory, is given by 
\begin{equation}
\label{eq:succ_prob_disorder}
P_w^{(1)}(t)=|\braket{w|\exp^{-iH_G t}|s}|^2\approx\frac{1}{1+n\epsilon_w^2/4}\sin^2(\frac{\Delta t}{2}).
\end{equation}
The probability $P^{(1)}_{w}(t)$ is maximum at $T'=\pi/\Delta$, $P^{(1)}_{w}(T')=1/(1+n\epsilon_w^2/4)$ and hence the algorithm needs to be repeated $1/P^{(1)}_{w}(T')$ times on average in order to find the marked vertex. This gives the average running time as
\begin{equation}
T_{dis}=\frac{\pi \sqrt{n}}{2}\sqrt{1+\frac{n \epsilon_w^2}{4}},
\end{equation}
where we assumed that $\epsilon_w$ takes the same value if one repeats the algorithm using the same system (it is a systematic error). We have thus two regimes of disorder:
\\~\\
\textbf{Weak diagonal disorder regime:} As long as $\sigma\leq\mathcal{O}(1/\sqrt{n})$, we have that $n\epsilon_w^2< 1$. Thus, in this regime the algorithm keeps an optimal running time of $\mathcal{O}(\sqrt{n})$ as after this time the probability of observing the solution state is a constant.\\~\\
\textbf{Strong diagonal disorder regime:} However beyond this threshold of $\sigma$, i.e.\ when $n\sigma^2\gg 1$, we expect that with high probability $n\epsilon_w^2\gg 1$ and thus the gap between the ground state and the first excited state is $\Delta\approx |\epsilon_w|/2\leq \sigma/2$. Also from Eq.~\eqref{eq:succ_prob_disorder} we find that after a time of $T_1=\pi/\epsilon_w=\mathcal{O}(\pi/\sigma)$, the probability of observing the solution is $\mathcal{O}(1/n\sigma^2)$. Thus the algorithm needs to be repeated $\mathcal{O}(n\sigma^2)$ times to obtain an average running time of $T=\mathcal{O}(n\sigma)$. If we assume that $\sigma$ depends on $n$ as $\sigma=n^ {-\alpha}$, the algorithm is sub-optimal for $\alpha<1/2$. The ground state of the Hamiltonian (not normalized) is given by 
\begin{equation}\label{eq:approx_ES}
\ket{\lambda_1^{(1)}}\approx
\begin{cases}
-\ket{w}+\frac{1}{|\epsilon_w|\sqrt{n}}\swbar+\mathcal{O}(\frac{1}{n\epsilon_w^2}), \text{~if~}\epsilon_w<0\\
-\frac{1}{|\epsilon_w|\sqrt{n}}\ket{w}+\swbar+\mathcal{O}(\frac{1}{n\epsilon_w^2}), \text{~if~}\epsilon_w>0
\end{cases}
\end{equation} 
and thus it has a large overlap either with $\ket{w}$ or with $\ket{s_{\bar{w}}}$ depending on the sign of the random variable $\epsilon_w$. As explained in the main text we can ensure that the ground state has always a larger overlap with the solution by shifting the parameter $\gamma$.\\

At this point, it is also important to understand the order of magnitude of the terms we have neglected in perturbation theory. The magnitude of the second order corrections to the eigenvalues $\lambda_1$ and $\lambda_2$ is of $\mathcal{O}(|V|^2)$. This means that we expect that Eq.~\eqref{eq:succ_prob_disorder} is valid for a timescale $t\ll 1/|V|^2$, which is sufficient for the discussion of the running time that we have done previously. Furthermore, it is possible to show that the terms we have neglected in the probability due to second order corrections to the eigenstates are of $\mathcal{O}(1/n)$.
\section{Bloch-Redfield Master Equation for a two level system}
\label{sec:bloch_redfield_two_level}
In this section we derive the Bloch-Redfield master equation describing the evolution of the system interacting with a thermal bath \cite{bookblum}. We assume that the system interacts with a thermal bath whose Hamiltonian is given by
\begin{equation} 
\label{eq:bath_hamiltonian}
H_{R}= \sum_{i=1}^n \sum_{\alpha} \omega_{\alpha} \adag_{i\alpha}a_{i\alpha},
\end{equation}
with $[a_{i\alpha},\adag_{j \beta}]=\delta_{i,j}\delta_{\alpha,\beta}$.
Furthermore, we will consider the interaction Hamiltonian given by 
\begin{equation}
\label{eq:interaction_hamiltonian}
H_I=\sum_{i=1}^n\sum_{\alpha}g_{i\alpha}(a_{i\alpha}+\adag_{i\alpha})\ket{i}\bra{i},
\end{equation}
i.e.\ each node of the graph is coupled to an independent bosonic bath, which we assume to be in a thermal state at temperature $1/\beta$ (throughout the article we are working in units where the Boltzmann constant $k_B=1$). 

The analog search Hamiltonian can be approximated by a two level system as long as the temperature of the bath is less than the gap between the first excited state and the rest of the energy levels. As seen previously, this is also true when the algorithm is affected by static disorder at the nodes of the graph. Throughout our analysis we shall assume that the bath temperature is such that the system can be well approximated by a two level system.

Let $\rho_{i j}=\bra{\lambda_i}\rho\ket{\lambda_j}$, where $\rho$ denotes the density matrix of the system. We are interested in 
calculating the time-evolution of the population of the solution which is given by
\begin{equation}
\label{eq:pop_at_solution}
P_{w}(t)=\rho_{11}(t)|\braket{w|\lambda_1}|^2+\rho_{22}(t)|\braket{w|\lambda_2}|^2+\rho_{12}(t)\braket{w|\lambda_1}\braket{\lambda_2|w}+\rho_{21}(t)\braket{w|\lambda_2}\braket{\lambda_1|w},
\end{equation}
where each $\rho_{ij}(t)$ is given by the solution to the Bloch-Redfield Master equation. That is
\begin{equation}
\label{eq:bloch_redfield_master_equation}
\dot{\rho_{ab}}=-i\omega_{ab}\rho_{ab}+\sum_{abcd}R_{abcd}\rho_{cd}(t),
\end{equation}
where $\omega_{ij}=\lambda_i-\lambda_j$. For a two level system, $\{a,b,c,d\}\in\{1,2\}$ and
\begin{equation}
\label{eq:bloch_redfield_tensor}
R_{abcd}=-\frac{1}{2}\sum_{j}\big{\{}\delta_{bd}\sum_{x}A^{j}_{ax}A^{j}_{xc}S_{j}(\omega_{cx})-A^{j}_{ac}A^{j}_{db}S_{j}(\omega_{ca})+\delta_{ac}\sum_{x}A^{j}_{dx}A^{j}_{xb}S_{j}(\omega_{dx})-A^{j}_{ac}A^{j}_{db}S_{j}(\omega_{db})\big{\}},
\end{equation}
such that $A^{j}_{xy}=c_{j x}c_{j y}^{*}$, where the coefficients $c_{ik}$ are obtained by writing the states $\ket{i}$ in the eigenbasis of the system as $\ket{i}=\sum_{k}c_{ik}\ket{\lambda_k}$. 
 Also,
\begin{align}
\label{eq:transition_rates}
S_{i}(\omega_{kl})=
\begin{cases}
 J(\omega_{kl})\mathcal{N}(\omega_{kl}),\text{~~~~~~~~~~~~~~~~~~} \omega_{kl}< 0\\
 J(\omega_{lk})\Big{(}\mathcal{N}(\omega_{lk})+1\Big{)},\text{~~~~~~~~~} \omega_{kl}\geq 0
\end{cases},
\end{align}
with $\mathcal{N}(\omega)=1/(e^{\beta\omega}-1)$ and $J(\omega)$ being the spectral density of the bath given by
\begin{equation}
\label{eq:bath_spectral_density}
J(\omega)=g^2\sum_{\alpha}\delta(\omega-\omega_\alpha),
\end{equation}
where it is assumed that the coupling between each site of the system and the bath is identical ($g_{i\alpha}=g,~\text{for all} ~i,\alpha$) and sufficiently weak so that the Markov approximation is valid. More precisely, the Markov approximation implies that the time scale of decay of the bath correlation functions $\delta t$ is much faster than the relevant time-scales of the system. We show in Appendix~\ref{appendix:markov_and_secular_approximation} that choosing $g\ll 1/\delta t$ ensures that the Markov approximation is valid. So in our analysis, we fix a value of $g$ that ensures the validity of this approximation. As the nodes the graph are coupled to a set of independent harmonic oscillators, each having the same spectral density we have that $S_{j}(\omega_{xy})$ is the same for all $j$. We drop this subscript henceforth. 

By expressing the two state system density matrix as 
\begin{align}
\label{eq:density_matrix_in_pauli_basis}
\rho=\dfrac{1}{2}\left(I+\vec{n}.\vec{\sigma}\right),
\end{align}
where $\vec{n}=(\rho_x,\rho_y,\rho_z)$ is a vector with real entries and $\sigma_{j}$'s are the Pauli matrices with $j\in\{x,y,z\}$. In the Pauli basis, the Bloch-Redfield master equation simplifies to the following set of differential equations:
\begin{align}
\label{eq:bloch_redfield_pauli_x}
\dot{\rho_x}&=-\omega_{12}\rho_y+S(\omega_{12})O_2\rho_z-\dfrac{S(0)}{2}O_3\rho_x\\
\label{eq:bloch_redfield_pauli_y}
\dot{\rho_y}&=\omega_{12}\rho_x-\left\{\dfrac{1}{2}S(0)O_3+O_1\left(S(\omega_{12})+S(\omega_{21})\right)\right\}\rho_y\\
\label{eq:bloch_redfield_pauli_z}
\dot{\rho_z}&=S(0)O_2 \rho_x-O_1\left(S(\omega_{12})+S(\omega_{21})\right)\rho_{z}+O_1\left(S(\omega_{21})-S(\omega_{12})\right),
\end{align}
where we have that
\begin{align}
\label{eq:system_operator_1}
O_1&=\sum_{i}(A^{i}_{12})^2\\
\label{eq:system_operator_2}
O_2&=\sum_{i}A^{i}_{12}(A^{i}_{11}-A^{i}_{22})\\
\label{eq:system_operator_3}
O_3&=\sum_{i}(A^{i}_{11}-A^{i}_{22})^2.
\end{align}
Throughout the article, we assume that the spectral density of the bath is ohmic with an exponential cut-off, i.e.\
\begin{equation}
J(\omega)=\eta g^2\omega e^{-\omega/\omega_c},
\end{equation}
where $\omega_c$ is the cut-off frequency of the bath and $\eta$ is a constant normalization factor. We fix the cut-off frequency $\omega_c$ to be a constant greater than one. For an ohmic bath, $S(0)=\lim_{\omega\rightarrow 0^{-}}S(\omega)=\lim_{\omega\rightarrow 0^{+}}S(\omega)=\eta g^2/\beta$. We shall use this general form of Bloch-Redfield master equation for analyzing how thermal relaxation can assist the analog search algorithm, even when static errors affect the algorithm.
\\ 
\section{Analog quantum search in the presence of a thermal bath}\label{appendix_C}
In this section we use the Bloch-Redfield equation derived previously to study the evolution of the system density matrix when there is no static disorder. In this scenario we have that the gap between the ground state and the first excited state $\omega_{21}=\Delta=2/\sqrt{n}$. Thus, to analyse whether the algorithm remains optimal, it is necessary to resolve the timesclae $\Delta^{-1}$ and so the secular approximation cannot be taken. Furthermore, we have that $\braket{w|\lambda_1}=\braket{w|\lambda_2}=1/\sqrt{2}$ and hence
\begin{equation}
P_{w}(t)=\frac{1}{2}\left(1+\rho_x(t)\right),
\end{equation}
where we used the fact that $\rho_{11}(t)+\rho_{22}(t)=1$ and that $2Re[\rho_{12}(t)]=\rho_x(t)$. 

To obtain the master equation corresponding to $\rho_{x}(t)$ observe that $A^{j}_{11}=A^{j}_{22}$. This implies that $O_2$ and $O_3$ in Eq.~\eqref{eq:system_operator_2} and Eq.~\eqref{eq:system_operator_3} are $0$. Furthermore,
$A^{i}_{21}=A^{i}_{12}=A_i$. This simplifies the Bloch-Redfield master equation considerably as $\rho_z$ is decoupled from $\rho_x$ and $\rho_y$. Thus to obtain the population of the solution state with time we have to solve the following differential equations:
\begin{align}
\label{eq:bloch_redfield_simplified_no_disorder}
\dot{\rho_x}&=\Delta\rho_y\\
\dot{\rho_y}&=-\Delta\rho_x-2\Gamma \rho_y,
\end{align}
where,
\begin{align}
\label{eq:transition_rate}
\Gamma&=\frac{1}{2}\sum_{i}A^2_{i}\left(S(\Delta)+S(-\Delta)\right)\\
&=\dfrac{(\sum_{i}A_i^2)J(\Delta)}{2\tanh(\beta\Delta/2)}\\
&=\dfrac{J(\Delta)}{8\tanh(\beta\Delta/2)}~~~~~~~~[~~\sum_{i}A_i^2=1/4+\mathcal{O}(1/n)~~].
\end{align}
The solution to the Bloch-Redfield Master Equation is
\begin{equation}
\label{eq:bloch_redfield_sigma_x}
\rho_{x}(t)=e^{-\Gamma t}\Big{[}\dfrac{\Gamma}{\sqrt{\Gamma^2-\Delta^2}}\Big{\{}\dfrac{e^{-\left(\sqrt{\Gamma^2-\Delta^2}\right)t}-e^{\left(\sqrt{\Gamma^2-\Delta^2}\right)t}}{2}\Big{\}}-\Big{\{}\dfrac{e^{-\left(\sqrt{\Gamma^2-\Delta^2}\right)t}+e^{\left(\sqrt{\Gamma^2-\Delta^2}\right)t}}{2}\Big{\}}\Big{]}.
\end{equation}
From Eq.~\eqref{eq:bloch_redfield_sigma_x} we find that there arise two distinct cases that determine the nature of relaxation dynamics:\\ 
~\\
(i)~Underdamped relaxation to the steady state $(\Gamma<\Delta)$: When $\sqrt{\Gamma^2-\Delta^2}$ is imaginary, we have that
\begin{equation}
\label{eq:pop_sol_under_damped_general}
P_{w}(t)=\dfrac{1}{2}\left(1-e^{-\Gamma t}\left\{\dfrac{\Gamma}{\Delta}\sin\left[\left(\sqrt{\Delta^2-\Gamma^2}\right)t\right]+\cos\left[\left(\sqrt{\Delta^2-\Gamma^2}\right)t\right]+\mathcal{O}(\Gamma^2/\Delta^2)\right\}\right),
\end{equation}
where 
\begin{equation}
\label{eq:rate_under_damped_general}
\Gamma=\mathcal{O}\left(\dfrac{\eta g^2\Delta}{\tanh(\beta\Delta/2)}\right).
\end{equation}
In this regime there is an oscillation timescale of $\mathcal{O}(\Delta^{-1})$ after which there is constant population at the solution.  Note that the relaxation timescale is longer than the oscillation timescale as the system reaches the steady state after a time $\mathcal{O}(1/\Gamma)$, with $P_w(\infty)=1/2$. Note that the larger the temperature the faster is the relaxation rate $\Gamma$ but the running time of the algorithm is still $\mathcal{O}(1/\Delta)$, since we are in the regime where $\Gamma<\Delta$.
~\\~\\
(ii)~Over-damped relaxation to the steady state ($\Gamma>\Delta$): In this case $\sqrt{\Gamma^2-\Delta^2}$ is real. Hence
\begin{equation}
\label{eq:pop_sol_over_damped_general}
P_{w}(t)=\dfrac{1}{2}(1-e^{-t\Delta^2/\Gamma})+\mathcal{O}(\Delta^2/\Gamma^2).
\end{equation}
Thus after a time $T=\mathcal{O}(\Gamma/\Delta^2)=\mathcal{O}(n\Gamma)$, the system reaches a steady state and the population of the solution is constant. Unlike the underdamped case, increasing the temperature makes the relaxation slower.

For a given system-environment coupling strength $g$, the parameter that determines whether we are in case (i) or (ii) is the temperature of the bath. In particular, we consider two regimes of temperature: the zero temperature case ($\beta\rightarrow\infty$) and the high temperature case ($\mathcal{O}(\log(n))\ll\beta\ll 1/\Delta$) where the temperature is higher than the energy of the first excited state but lower than the energy of the higher excited states. We do not analyse the intermediate case when $1/\Delta\leq\beta <\infty$ as the bath correlation time, given by $\mathcal{O}(1/\beta)$ (see Sec.~V and VII), becomes larger than the system time scale of $1/\Delta$ and thus the Markovian approximation is not valid.
\\~\\
\textbf{Zero temperature ($\beta\rightarrow\infty$):} When the thermal bath is at near zero temperature we have that
\begin{align}
\label{eq:transition_rate_no_disorder_zero_temp_1}
\Gamma&=\mathcal{O}\left(\eta g^2\Delta \right)\\
\label{eq:transition_rate_no_disorder_zero_temp_2}
      &=\mathcal{O}\left(\dfrac{\eta g^2}{\sqrt{n}} \right).
\end{align}

Since $g\ll 1$, we are always in the underdamped regime when the thermal bath is at zero temperature and the population of the solution state is given by Eq.~\eqref{eq:pop_sol_under_damped_general}. So the system oscillates with a period of $\mathcal{O}(\sqrt{n})$ and the probability of being at the solution is a constant which gives the optimal scaling of the running time of the analog search algorithm. After a timescale of $\Gamma^{-1}=\mathcal{O}(\sqrt{n}/g^2)$ these oscillations are damped and the system converges to the steady state and hence the algorithm exhibits a fixed point behavior. 
\\~\\
\textbf{High temperature regime ($\mathcal{O}(\log(n))\ll\beta\ll 1/\Delta$):} In this regime, we consider the scenario where the temperature of the bath is greater than $\Delta$, but sufficiently low ($\beta>>\log(n)$) to ensure that the two level approximation is valid. 
Note that in this scenario, $\tanh(\beta\Delta/2)\approx\beta\Delta/2$. As discussed previously and shown in Sec.~VII, the correlation time of the thermal bath, $\delta t=\mathcal{O}(\beta)$. The rate $\Gamma$ is 
\begin{align}
\label{eq:transition_rate_no_disorder_high_temp}
\Gamma&=\dfrac{\eta g^2\Delta e^{-\Delta/\omega_c}}{8\tanh(\beta\Delta/2)}\\
      &=\mathcal{O}\left(\dfrac{\eta g^2}{\beta}\right).
\end{align}
In this case, whenever $g^2/\beta\leq \mathcal{O}(1/\sqrt{n})$, we are in the underdamped regime and the algorithmic running time is optimal. Otherwise, the system is in the overdamped regime and the relaxation time is
\begin{equation}
T_{rel}(\beta)=\mathcal{O}\left(\dfrac{\eta g^2 n}{\beta}\right).
\end{equation}
In this regime the relaxation time gets slower with increase in temperature and an advantage with respect to the classical running time for search is only possible if the ratio $g^2/\beta$ decreases with $n$.
\section{Analog quantum search with diagonal disorder in the presence of a thermal bath}\label{appendix_D}
In this section, we study the dynamics of the search algorithm coupled to a thermal bath and in the presence of static disorder using the Bloch-Redfield equation. This equation allows us to resolve timescales of the order of the inverse of system gap $\Delta^{-1}\approx (\epsilon_w^2+4/n)^{-1/2}$, as the secular approximation is not taken. This is particularly important in the weak disorder regime, which is not treated in the main text. In this regime, we have $\Delta^{-1}=\mathcal{O}(\sqrt{n})$ and hence, being able to resolve such timescales is important to understand whether the search algorithm runs in optimal time.  

As it can be seen from the analysis in Appendix~\ref{sec:analog_search_static_errors}, the presence of static disorder changes the ground state and the first excited state of the algorithm. In fact now $|\braket{\lambda_1|w}|\neq |\braket{\lambda_2|w}|$. In the presence of a thermal bath, we require that the ground state of the system Hamiltonian has a higher overlap with the solution state in order to enhance the population at the solution via thermal relaxation. To ensure that the ground state of the search Hamiltonian has a higher overlap with the solution state $\ket{w}$ we need $\braket{w|H_{red}|w}<\braket{s_{\bar{w}}|H_{red}|s_{\bar{w}}}$, which can be achieved by an appropriate choice of the parameter $\gamma$, as discussed in the main text. A possible choice is $\gamma=(1-\sigma)/n$. This choice ensures that when the thermal bath is at low temperatures, the success probability of the algorithm is higher, although the relaxation time is slower. For example, when the thermal bath is at zero temperature, this choice of $\gamma$ ensures that the system relaxes to the solution state.

In this case, the gap between the ground state and the first excited state, as result of this choice of $\gamma$ is 
\begin{equation}
\label{eq:gap_disorder}
\Delta=\sigma-\epsilon_w+\mathcal{O}(1/\sqrt{n}).
\end{equation}
As typically $\epsilon_w=\mathcal{O}(\sigma)$, we have that $\Delta=\mathcal{O}(\sigma)$. Also from Sec.~I, there are two regimes of static disorder and for each of which the analysis for the relaxation of the system is going to differ. 
\subsection*{Weak diagonal disorder} In this regime, i.e.\ when the strength of disorder, $\sigma<\mathcal{O}(1/\sqrt{n})$, the analog search algorithm remains robust to this error and the optimal running time is maintained. Note that the ground state and the first excited state have a constant overlap with the solution state. As mentioned previously, the new choice of $\gamma$ ensures that the ground state has a higher overlap with the solution state as compared to the first excited state. Also the gap between the ground state and the first excited state $\Delta\sim\mathcal{O}(1/\sqrt{n})$.

\begin{figure}[h!]
        \centering
        \includegraphics[scale=0.7]{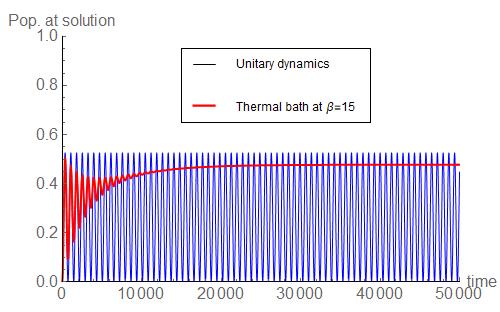}
          \caption{\small{Comparison of population at the solution node with time for a complete graph of $100000$ nodes where each node of the graph is affected by weak diagonal disorder of standard deviation $\sigma=0.006$ in the unitary regime and in the presence of a thermal bath having a cut-off frequency of $\omega_c=2$ and system-bath coupling $g=0.04$. The oscillatory thin blue curve indicates the population in the unitary scenario, i.e.\ in the absence of a thermal bath. The thick red curve shows the population at the solution in the presence of a thermal bath at inverse temperature, $\beta=15$. The steady state of the thermal bath has an overlap of close to $1/2$ with the solution state.}}
       \label{fig:weak_dis_nonzero_temp}
\end{figure}

In the presence of the thermal bath, the behavior of the analog search algorithm is similar to the scenario where there was no disorder. However, in this regime $A^{i}_{11}\neq A^{i}_{22}$ and so $O_2$ and $O_3$ are non-zero. The Bloch-Redfield equations in Eq.~\eqref{eq:bloch_redfield_pauli_x}-\eqref{eq:bloch_redfield_pauli_z} are written as 
\begin{equation}\label{disordered_BReq}
\begin{bmatrix}
\dot{\rho_x}\\
\dot{\rho_y}\\
\dot{\rho_z}
\end{bmatrix}
=
\underbrace{\begin{bmatrix}
-\frac{1}{2}S(0)O_3 && -\omega_{12} && S(\omega_{12})O_2\\
\omega_{12} && -\frac{1}{2}S(0)O_3-2\Gamma && 0\\
S(0)O_2 && 0 && -2\Gamma
\end{bmatrix}}_M
\begin{bmatrix}
\rho_x\\
\rho_y\\
\rho_z
\end{bmatrix}
+
\begin{bmatrix}
0\\
0\\
O_1(S(\omega_{21})-S(\omega_{12}))
\end{bmatrix},
\end{equation}
where $\omega_{21}=\Delta$ and $\Gamma=O_1 J(\Delta)\coth(\beta\Delta/2)/2$. The quantities $O_1,O_2$ and $O_3$ are $\mathcal{O}(1)$.
\\~\\
\textbf{Zero temperature ($\beta \rightarrow \infty$):} At zero temperature, $S(0)=g^2/\beta=0$ which simplifies the master equation and thus the calculation of the eigenvalues and eigenvectors of the matrix $M$ of Eq.~\eqref{disordered_BReq}. We find that $\Gamma=\mathcal{O}\left(\eta g^2/\sqrt{n}\right)$ and the system reaches the steady state after a time of $\mathcal{O}(\sqrt{n}/g^2)$ which has the same scaling as the case where no static error is present (See Eq.~\eqref{eq:transition_rate_no_disorder_zero_temp_2}). 
\\~\\
\textbf{High temperature ($\mathcal{O}(\log(n))\ll\beta\ll 1/\Delta$):} In this regime, no simplification to the master equation is possible and we resort to numerical simulations. Intuitively, one would expect that the  behavior of the algorithm is similar to the scenario where there was no disorder. We numerically verify that this is indeed the case and plot the population at the solution with time at high temperature in Fig.~\ref{fig:weak_dis_nonzero_temp}. We observe that the probability of success oscillates for small times and eventually the system relaxes to the steady state which is expected to be a statistical mixture between the solution state $\ket{w}$ and the equal superposition of the rest of the nodes ($\swbar$).
\subsection*{Strong diagonal disorder} 
When the strength of disorder $\sigma>1/\sqrt{n}$, the analog search algorithm loses its optimality. In the unitary case, one observes that one needs to measure after a time $T=\mathcal{O}(\pi/\sigma)$, to find the solution with probability $\mathcal{O}(1/(n\sigma^2))$. Furthermore, this probability is amplified by repeating the algorithm $\mathcal{O}(n\sigma^2)$ times, thereby obtaining an expected running time of $T=\mathcal{O}(n\sigma)$. We show that the introduction of a thermal bath can amplify the amplitude of the solution node. In fact, increasing the temperature ensures faster relaxation to the steady state which has a high overlap with the solution state. Moreover, the resultant dissipative dynamics ensures that the population at the solution node only increases with time thereby circumventing the need to repeat the algorithm several times as in the unitary case. 

Choosing $\gamma=(1-\sigma)/n$ yields the approximate eigenstates (not normalized) as
\begin{align}
\label{eq:approx_eigenstates_disorder1}
\ket{\lambda_{1}}&\approx \ket{w}+\frac{1}{\sqrt{n} (\sigma-\epsilon_w)}\ket{s_{\bar{w}}},\\
\label{eq:approx_eigenstates_disorder2}
\ket{\lambda_{2}}&\approx \frac{1}{\sqrt{n} (\sigma-\epsilon_w)}\ket{w}-\ket{s_{\bar{w}}},
\end{align}
by neglecting terms higher order terms that will remain small as long as $\sigma\ll 1$, where the gap $\lambda_2-\lambda_1$ is 
\begin{equation}
\label{eq:gap_strong_disorder}
\Delta\approx \sigma-\epsilon_w=\mathcal{O}(\sigma).
\end{equation}
Thus there is a gap of $1-\Delta=1-\mathcal{O}(\sigma)$ between the first excited state and the rest of the spectrum which is a constant as long as $\sigma\ll 1$. This enables us to approximate our system as a two level system for low temperatures, i.e.\ $\beta\gg\mathcal{O}(\log(n))$.

The population at the solution, given by Eq.~\eqref{eq:pop_at_solution} is now
\begin{equation}
\label{eq:pop_at_solution_general_strong_disorder}
P_w(t)\approx\rho_{11}(t)+\mathcal{O}\left(\dfrac{1}{\sigma\sqrt{n}}\right),
\end{equation}
which implies that the population of the solution is determined by the population of the ground state for $1/\sqrt{n}\ll\sigma\ll 1$. 

In this regime of static disorder, we can coarse grain the time-scale of the relaxation of the system which simplifies the Bloch-Redifield equation considerably. Note that the Bloch-Redfield equation already assumes a coarse graining in the time-scale of the system owing to the Markov approximation.  Furthermore the time-scales that we are interested in ($\sim \sqrt{n}$) is significantly greater than the gap $\Delta=\mathcal{O}(\sigma)$, we can take  the so-called secular approximation which implies an additional course graining in the relaxation dynamics of the system. In general, if $g$ is the strength of coupling between the system and the bath and $\delta t$ is the width of the correlation function of the bath, the typical relaxation time-scale of the system is $\sim 1/(g^2\delta t)$ and for the secular approximation to hold this has to be greater than $1/\Delta$. Thus, we fix a $g$ that respects both the secular and the Markov approximation. Whenever $\beta\sigma\ll 1$, the choice of $g$ that respects the secular approximation, also respects the Markov approximation. For further details refer to the Sec.~VII.

Henceforth, in this section we shall assume that $g$ is such that in addition to the Markov approximation, the secular approximation also holds. Taking the secular approximation ensures that in the Bloch-Redfield equation, the diagonal terms of the density matrix never couples with the off-diagonal terms (Lindblad form). Since from Eq.~\eqref{eq:pop_at_solution_general_strong_disorder}, we find that the population of the ground state determines the population of the solution, we have the master equation of the dynamics of the population of the ground state and the first excited state. 
\begin{align}
\label{eq:master_eq_pop_secular}
\dot{\rho}_{kk}&= \sum_{l\neq k}W_{kl}\rho_{ll}-\sum_{l\neq k}W_{lk}\rho_{kk},
\end{align}
where $k\in\{1,2\}$ and $l\in\{1,2,\dots,n\}$. The transition rates are given by
\begin{align}
\label{eq:transition_rates_pop_secular}
W_{kl}=\begin{cases}
2\pi J(\omega_{kl})\Lambda_{kl}\mathcal{N}(\omega_{kl}),\text{~~~~~~~~~~~~~~~~~~} \omega_{kl}< 0\\
2\pi J(\omega_{lk})\Lambda_{kl}\Big{(}\mathcal{N}(\omega_{lk})+1\Big{)},\text{~~~~~~~~~} \omega_{kl}\geq 0
\end{cases}
\end{align} 
such that $\Lambda_{kl}=\sum_{i}|c_{ik}c_{il}|^2$. Solving the differential equation \eqref{eq:master_eq_pop_secular} we obtain,
\begin{align}
\label{sol_diff_eq}
\rho_{11}&=\frac{W_{12}}{W_{12}+W_{21}}\left(1-e^{-(W_{12}+W_{21})t}\right),\\
&=\frac{1}{1+e^{-\beta\Delta}}\left(1-e^{-t/T_{rel}}\right)+\frac{e^{-t/T_{rel}}}{n}
\end{align}
with the relaxation time given by 
\begin{equation}
T_{rel}=\frac{1}{W_{12}+W_{21}}.
\end{equation}
On substituting the appropriate terms we obtain
\begin{equation}
\label{eq:relaxation_time_general_expression}
T_{rel}\sim\dfrac{1}{\Lambda_{12} J(\Delta)}\tanh\left(\dfrac{\beta\Delta}{2}\right).
\end{equation}
\section{Validity of the Markov and secular approximations}
\label{appendix:markov_and_secular_approximation}
In this section we discuss for what regime of system-bath coupling parameters the Markov and secular aproximations are expected to hold \cite{bookblum, albash2012quantum}. Let $H_S$ represent the Hamiltonian of the system while $H_E$ be the Hamiltonian of the environment. Consider the following interaction Hamiltonian
\begin{equation}
\label{eq:interaction_hamiltonian_example}
H_I=g\sum_i Q_iF_i,
\end{equation}
where $Q_i$ are operators acting on the system's Hilbert space and $F_i$ are operators acting on the Hilbert space of the bath. So in the interaction picture let
\begin{equation}
\label{eq:interaction_hamiltonian_example2}
H_I(t)=g\sum_i Q_i(t)F_i(t),
\end{equation}
where $Q_i(t)$ are and $F_i(t)$ are the previously defined operators in the interaction picture. 
Thus we obtain that, after tracing out the environment degrees of freedom, the evolution of the reduced density matrix of the system is given by the Bloch-Redfield master equation which is of the following form:
\begin{equation}
\frac{d\rho_{S}}{dt}=\left(\mathcal{L}_{unitary}+\mathcal{L}_{diss}\right)\rho_S(t),
\end{equation}
where $\mathcal{L}_{unitary}$ is the super-operator corresponding to purely unitary dynamics while $\mathcal{L}_{diss}$ corresponds to the super-operator corresponding to purely dissipative dynamics.

The Born approximation is respected as long as we are in the weak coupling regime, i.e. $||\mathcal{L}_{relax}||\ll||\mathcal{L}_{unitary}||=1$.
Note that in the interaction picture, the dynamics of the reduced density matrix of the system (up to $\mathcal{O}(g^2)$) is given by
\begin{equation}
\frac{d\rho_{SI}}{dt}=g^2\sum_{ij}\int_{0}^{\infty} dt' ~Q_i(t')\rho(t')_{SI}Q_j(t')F_{ij}(t')+\text{~other similar terms}
\end{equation}
where $F_{ij}(t)=\langle F_i(t) F_j\rangle$ is the bath correlation function. The bath correlation function decays after bath correlation time-scale defined in the article as $\delta t$. Thus the term inside the integral, i.e. $||\mathcal{L}_{diss}||=\mathcal{O}(g^2\delta t)$. Now for the Markov approximation to be valid we require that the bath correlation decays faster than the typical time scale of relaxation of the system. This implies that 
\begin{align}
\label{eq:weak_coupling_markov_approximation}
&\delta t<<\dfrac{1}{g^2\delta t}\\
\implies & g<<\frac{1}{\delta t}.
\end{align}
In the main text we analyse the analog search algorithm affected by strong diagonal disorder, using a master equation where the secular approximation was used. This means that the typical timescale of relaxation of the system due to the coupling with the bath should be greater than typical system timescale given by the inverse of the gap $\Delta$ between ground and first excited states. Hence, we should have
\begin{align}
\label{eq:weak_coupling_secular_approximation}
&\frac{1}{\Delta}<<\dfrac{1}{g^2\delta t}\\
\implies & g<<\sqrt{\frac{\Delta}{\delta t}}.
\end{align}
Then, for both secular and Markov approximations to be respected, we need the value of the coupling strength $g<min\{1/\delta t,\sqrt{\Delta/\delta t}\}$.
\section{Correlation function of an ohmic bath with an exponential cutoff at zero temperature}
\label{sec:bath_correlation_function_zero_temp}
In this section we calculate the width of the bath correlation function at $0$ temperature \cite{bookblum} which is an important quantity to understand the regime of validity of the Markov and secular approximations (see Sec.~V). The spectral density of this bath is given by
\begin{equation}
\label{linear_coupled_bath}
J(\omega)=\eta g^2\omega e^{-\omega/\omega_c},
\end{equation}
where $\omega_c$ is the bath cut off frequency and if $0<d<1$, the bath is sub-ohmic, for $d=1$, the bath is ohmic while for $d>1$, the bath is super-ohmic. 
On the other hand, the bath correlation function is given by
\begin{equation}
\label{bath_correlation}
F_{ii}(t)=\langle F_i(t) F_i\rangle=\int_{0}^{\infty} J(\omega)\Big{(}\coth(\beta\omega/2)\cos(\omega t)-i\sin(\omega t)\Big{)} d\omega,
\end{equation}
where $F_i$ and $F_i(t)$ are defined in Eq.~\eqref{eq:interaction_hamiltonian_example} and Eq.~\eqref{eq:interaction_hamiltonian_example2} respectively. Also $\langle O \rangle$ represents the expectation value of operator $O$.

For the Markovian approximation to be valid, the width of the correlation function should decay much faster than the relevant timescales of the system.\\                                                                                                                                                                                                                                                                                                                                                                                                                                                                                                                                                                                                                                                                                                                                                                                                                                                                                                                                                                                                                                                                                                                                                                                                                                                                                                                                                                                                                                                                                                                                                                                                                                                                                                                                                                                                                                                                                                                                                                                                                                                                                                                                                                                                                                                                                                                                                                                                                                                                                                                                                                                                                                                                                                                                                                                                                                                                                                                                                                                                                                                                                                                                                                                   
At zero temperature,
\begin{align}
\label{bath_corr_ohmic}
F_{ii}(t)&=\eta g^2\int_{0}^{\infty}\omega e^{-\omega/\omega_c} e^{-i\omega t} d\omega\\
         &=\eta g^2 \int_{0}^{\infty}\omega e^{-\omega(\frac{1}{\omega_c}+it)} d\omega\\
         &=\frac{\eta g^2\omega_c^2}{(1+it\omega_c)^{2}}\int_{0}^{\infty}q e^{-q} dq\\
         &\Big{[}\text{Considering~}q=\omega\Big{(}\frac{1}{\omega_c}+it\Big{)}\Big{]}\\
         &=\frac{\eta g^2\omega_c^2}{(1+it\omega_c)^{2}}.
\end{align}
So the width of $F_{ii}(t)$ is $\delta t=\mathcal{O}(1/\omega_c)$.
\section{Correlation function of an ohmic bath with an exponential cutoff at non-zero temperatures}
\label{appendix:bath_correlation_function_non_zero_temp}
In this section we calculate the width of the bath correlation function at finite temperature which is an important quantity to understand the regime of validity of the Markov and secular approximations (see Sec.~V). We follow arguments similar to that of Ref.~\cite{albash2012quantum}.

We consider baths with an ohmic spectral density as in Eq.~\eqref{linear_coupled_bath}. Considering the bath correlation function defined in Eq.~\eqref{bath_correlation} we have that 
\begin{align}
\label{eq:bath_correlation_finite_temperature}
F_{ii}(t)&=\int_{0}^{\infty} J(\omega)\Big{\{}\Big{[}\frac{1+e^{-\beta\omega}}{1-e^{-\beta\omega}}\Big{]}\Big{(}\frac{e^{i\omega t}+e^{-i\omega t}}{2}\Big{)}-
\Big{(}\frac{e^{i\omega t}-e^{-i\omega t}}{2}\Big{)}\Big{\}}d\omega\\
&=\int_{0}^{\infty}\frac{J(\omega)}{2(1-e^{-\beta\omega})}\Big{(}e^{-i\omega t}+
e^{i\omega t-\beta\omega}\Big{)}d\omega\\
&=\omega_c^{1-d}\Big{[}\underbrace{\int_{0}^{\infty} \frac{\omega^d}{1-e^{-\beta\omega}}e^{-\omega (it+1/\omega_c)}d\omega}_{\mathcal{I}_1}+
\underbrace{\int_{0}^{\infty} \frac{\omega^d}{1-e^{-\beta\omega}}e^{-\omega( -it+\beta+1/\omega_c)}d\omega}_{\mathcal{I}_2}\Big{]}.
\end{align}
First we consider the integral $\mathcal{I}_1$. We have
\begin{align}
\label{eq:I1}
\mathcal{I}_1&=\int_{0}^{\infty} \frac{\omega^d}{1-e^{-\beta\omega}}e^{-\omega (it+1/\omega_c)}d\omega\\
&=\frac{1}{\beta^{d+1}}\int_{0}^{\infty}\frac{q^d e^{-qz}}{1-e^{-q}}dq \text{~~~~~~~~~~~~~~~~}[q=\beta\omega\text{~ and~}z=\frac{it}{\beta}+\frac{1}{\beta\omega_c}]\\
&=\frac{(-1)^{d+1}}{\beta^{d+1}}\psi^{(d)}(z),
\end{align}
where $\psi^{(d)}(z)$ is the polygamma function defined as $\psi^{n}(z)=\dfrac{d^{m+1}}{dz^{m+1}}\ln\Gamma(z)$, where $\Gamma(z)=\int_{0}^{\infty}e^{-x}x^{z-1} dx$ is the Gamma function. So $\psi^{n}(z)=\int_{0}^{\infty}\frac{q^n e^{-qz}}{1-e^{-q}}dq$.

Following similar arguments we have that 
\begin{equation}
\label{eq:I2}
\mathcal{I}_2=\frac{(-1)^{d+1}}{\beta^{d+1}}\psi^{(d)}(1+\frac{1}{\beta\omega_c}-\frac{it}{\beta}).
\end{equation}
Thus the bath correlation function is 
\begin{equation}
\label{eq:bath_correlation_finite_temperature_final}
F_{ii}(t)=\frac{(-1)^{d+1}\eta g^2\omega_c^{1-d}}{\beta^{d+1}}\Big{[}\psi^{(d)}(\frac{1}{\beta\omega_c}+\frac{it}{\beta})+\psi^{(d)}(1+\frac{1}{\beta\omega_c}-\frac{it}{\beta})\Big{]}.
\end{equation}
We shall assume that the quantity $\beta\omega_c>>1$ and expand the polygamma functions in Eq.~\eqref{eq:bath_correlation_finite_temperature_final} according to Taylor series. Firstly, observe that $\frac{d^{m}}{dz^m}\psi^{n}(z)=\psi^{m+n}(z)$. Then,
\begin{align}
\label{eq:first_term_taylor_polygamma}
&\psi^{(d)}\left(\frac{1}{\beta\omega_c}+\frac{it}{\beta}\right)=\psi^{(d)}\left(\frac{it}{\beta}\right)+\sum_{n=1}^{\infty}\dfrac{\psi^{(n+d)}\left(\frac{it}{\beta}\right)}{(\beta\omega_c)^n n!}\text{~and},\\
\label{eq:second_term_taylor_polygamma}
&\psi^{(d)}\left(1+\frac{1}{\beta\omega_c}-\frac{it}{\beta}\right)=\psi^{(d)}\left(1-\frac{it}{\beta}\right)+\sum_{n=1}^{\infty}\dfrac{\psi^{(n+d)}\left(1-\frac{it}{\beta}\right)}{(\beta\omega_c)^n n!}.
\end{align}
For simplicity, henceforth we shall concern ourselves with the case where the bath is ohmic $(d=1)$ and make statements for $d$ in general at the end. Thus combining Eqs.~\eqref{eq:bath_correlation_finite_temperature_final}, \eqref{eq:first_term_taylor_polygamma} and \eqref{eq:second_term_taylor_polygamma} we have that the bath correlation function is
\begin{equation}
\label{eq:bath_correlation_finite_temperature_simplified}
F_{ii}(t)=\frac{\eta g^2}{\beta^2}\left[
\psi^{(1)}\left(\frac{it}{\beta}\right)+
\psi^{(1)}\left(1-\frac{it}{\beta}\right)+
\sum_{n=1}^{\infty}\dfrac{\psi^{(n+1)}\left(\frac{it}{\beta}\right)+
\psi^{(n+1)}\left(1-\frac{it}{\beta}\right)}{(\beta\omega_c)^n n!}\right].
\end{equation}
Now we shall simplify Eq.~\eqref{eq:bath_correlation_finite_temperature_simplified} using a couple of properties of polygamma functions. Let us state these two properties first.
\begin{align}
\label{eq:reflection_relation_polygamma}
&\psi^{(n)}(1-z)+(-1)^{(n-1)}\psi^{(n)}(z)=(-1)^n\pi\dfrac{d^n}{dz^n}\cot(\pi z)\\
\label{eq:reccurence_relation_polygamma}
&\psi^{(n)}(z+1)=\psi^{(n)}(z)+\dfrac{(-1)^n n!}{z^{n+1}}.
\end{align}
Using Eq.~\eqref{eq:reflection_relation_polygamma} for $n=1$ and $z=it/\beta$ we have that
\begin{equation}
\label{eq:polygamma_solution_using_reflection}
\psi^{(1)}\left(\frac{it}{\beta}\right)+
\psi^{(1)}\left(1-\frac{it}{\beta}\right)=-\pi^2\csch^2(\pi t/\beta).
\end{equation}
Also using Eq.~\eqref{eq:reccurence_relation_polygamma}, we have that
\begin{equation}
\label{eq:polygamma_solution_using_recurrence}
\psi^{(n+1)}\left(1-\frac{it}{\beta}\right)=\psi^{(n+1)}\left(\frac{-it}{\beta}\right)+\dfrac{(-1)^n n!}{{(-it/\beta)}^{n+2}}.
\end{equation}
Substituting the results of Eq.~\eqref{eq:polygamma_solution_using_reflection} and Eq.~\eqref{eq:polygamma_solution_using_recurrence} into Eq.~\eqref{eq:bath_correlation_finite_temperature_simplified} we obtain
\begin{align}
\label{eq:bath_correlation_finite_temperature_polygamma_substituted_1}
F_{ii}(t)&=\frac{\eta g^2}{\beta^2}\left[
-\pi^2\csch^2(\pi t/\beta)
+
\sum_{n=1}^{\infty}\dfrac{(-1)^{n+1} (n+1)!}{{(-it/\beta)}^{n+2}(\beta\omega_c)^n n!}
+
\sum_{n=1}^{\infty}\dfrac{\psi^{(n+1)}\left(\frac{it}{\beta}\right)+
\psi^{(n+1)}\left(-\frac{it}{\beta}\right)}{(\beta\omega_c)^n n!}\right]\\
\label{eq:bath_correlation_finite_temperature_polygamma_substituted_2}
&= \eta g^2\left[
-\frac{\pi^2}{\beta^2}\csch^2(\pi t/\beta)
+
\sum_{n=1}^{\infty}\dfrac{(-1)^{n+1} (n+1)}{{(-it)}^{n+2}(\omega_c)^n}
+
\sum_{n=1}^{\infty}\dfrac{\psi^{(n+1)}\left(\frac{it}{\beta}\right)+
\psi^{(n+1)}\left(-\frac{it}{\beta}\right)}{(\beta\omega_c)^n n!}\right].
\end{align}
From Eq.~\eqref{eq:bath_correlation_finite_temperature_polygamma_substituted_2}, we find that the bath correlation time depends on both $\beta$ and $\omega_c$. Assume that $\omega_c>1$ and that we are interested in time-scales that are larger than the thermal time-scale (i.e.\ $t>>\beta$) implying that $\csch^2(\pi t/\beta)\approx e^{-2\pi t/\beta}+\mathcal{O}(e^{-4\pi t/\beta})$. In this regime the bath correlation time-scales are $\delta t\sim \mathcal{O}(\beta)$.
\section{Lower bound on the optimality of analog quantum search in the presence of an environment}
\label{sec:lower_bound_search_environment}
We prove that the running time of the analog quantum search algorithm is lower bounded by $\mathcal{O}(\sqrt{n})$ in the presence of an environment of arbitrary dimension. Our derivation also shows that the running time of this algorithm cannot be improved any further by appending an ancillary space to the original search space. We follow an argument that is similar to Ref.~\cite{Farhi_analog_grover}.  

We are given an oracular Hamiltonian, $H_w$ that marks the search node and add to it a time dependent drive Hamiltonian $H_D(t)$.  Let us assume that an ancillary space of dimension $M$ is appended to the search space (in this case of dimension $n$). In such a case, the oracle Hamiltonian is
\begin{equation}
\label{eq:search_oracle_proof}
H_w=\ket{w}\bra{w}\otimes I_M,
\end{equation}
where $I_M$ is the identity matrix of dimension $M$. This implies that the oracle marks a node in the search space alone. If the basis states of the environment are $\{\ket{j}\}$ for $1\leq j\leq M$, then
\begin{align}
\label{eq:search_oracle_extended}
H_w&=\ket{w}\bra{w}\otimes \Big{(}\sum_{j=1}^{M}\ket{j}\bra{j}\Big{)}\\
&=\sum_{j=1}^{M}\ket{w}\bra{w}\otimes \ket{j}\bra{j}.
\end{align}
Notice that the oracle is in fact marking $M$ elements in the Hilbert space spanned by the system and the environment of dimension $nM$. Also 
\begin{equation}
\sum_{w}H_w=I_{nM}
\end{equation}
is the sum of $(nM)/M=n$  number of disjoint possible marked states in the total $nM$-dimensional Hilbert space.

The driver Hamiltonian $H_D(t)$ acts on the total Hilbert space. Thus the total search Hamiltonian is given by
\begin{equation}
\label{eq:total_search_hamiltonian_proof_optimality}
H_{search}=H_w+H_D(t).
\end{equation}
This formalism is enough to capture the scenarios where the system under consideration (the underlying graph) undergoes interactions with the environment. The driver Hamiltonian encompasses both the Hamiltonian of the environment, the interaction Hamiltonian as well as the system Hamiltonian proportional to the graph's adjacency matrix . Assume that the initial state of the algorithm is in some pure state $\ket{\psi_0}\in \mathbb{C}^{nM}$. If the state $\ket{w}$ is marked, let us assume that after a time $t$ we obtain the algorithm is in state $\ket{\psi_w(t)}$. Now if a different state was marked, say $\ket{w'}$, and the algorithm commenced from the same initial state $\ket{\psi_0}$, then in order to ensure sufficient distinguishability between $\ket{w}$ and $\ket{w'}$, the states $\ket{\psi_w(t)}$ and $\ket{\psi_{w'}(t)}$ should be sufficiently distinguishable. In fact for this to happen $\ket{\psi_w(t)}$ should be sufficiently different from any $\ket{w}$-independent state $\ket{\psi(t)}$ resulting from the evolution of $\ket{\psi_0}$ under the Hamiltonian $H_D(t)$. We want to ensure that, for any $w$, after some large enough time $T$ we have
\begin{equation}
\label{eq:separation_between_states}
||\ket{\psi_w(T)}-\ket{\psi(T)}||^2\geq \epsilon,
\end{equation}
which implies that
\begin{equation}
\label{eq:lower_bound_separation_optimality_proof}
\sum_w ||\ket{\psi_w(T)}-\ket{\psi(T)}||^2\geq n\epsilon.
\end{equation}
Now we intend to obtain an upper bound for the rate of change in the norm squared of the separation between the aforementioned states, i.e.\
\begin{align}
\label{eq:derivative_separation_between_states_over_solutions}
\dfrac{d}{dt}||\ket{\psi_w(t)}-\ket{\psi(t)}||^2&=-2~\text{Re}\dfrac{d}{dt}\braket{\psi_w(t)|\psi(t)}\\
                                                &=2~\text{Im}\braket{\psi_w(t)|H_w|\psi(t)}\\
                                                &\leq 2||H_w\ket{\psi(t)}||.
\end{align}
Thus
\begin{equation}
\dfrac{d}{dt}\sum_w||\ket{\psi_w(t)}-\ket{\psi(t)}||^2\leq 2 \sum_w ||H_w\ket{\psi(t)}||.
\end{equation}
Now let 
\begin{equation}
\ket{\psi(t)}=\sum_{i=1}^{n}\sum_{j=1}^{M} a_{ij}\ket{i}\ket{j},
\end{equation}
where $\sum_{i=1}^{n}\sum_{j=1}^{M} |a_{ij}|^2=1$. Thus
\begin{equation}
H_w\ket{\psi(t)}=\sum_{j=1}^{M}a_{wj}\ket{w}\ket{j}.
\end{equation}
Let $p_w=\sum_{j=1}^{M}|a_{wj}|^2\leq 1$. Since $\sum_{w=1}^n p_w=1$ it implies that $\sum_{w=1}^n \sqrt{p_w}\leq \sqrt{n}$. 
Thus we have that 
\begin{align}
\dfrac{d}{dt}\sum_w||\ket{\psi_w(t)}-\ket{\psi(t)}||^2&\leq 2\sum_w||H_w\ket{\psi(t)}||=2\sum_{w=1}^n \sqrt{p_w}\leq 2\sqrt{n}.
\end{align}
This gives the following upper bound:
\begin{equation}
\label{eq:upper_bound_separation_optimality_proof}
\sum_w||\ket{\psi_w(T)}-\ket{\psi(T)}||^2\leq 2\sqrt{n}T.
\end{equation}
Combining Eq.~\eqref{eq:lower_bound_separation_optimality_proof} and Eq.~\eqref{eq:upper_bound_separation_optimality_proof}, we obtain that
\begin{equation}
\label{eq:optimality_proof_final_time_single_sol}
T\geq \frac{\sqrt{n}\epsilon}{2}.
\end{equation}
\bibliographystyle{unsrt}
\bibliography{Bibliography}
\end{document}